\documentclass[aps,prl,superscriptaddress,twocolumn]{revtex4} 

\begin{document}

\title{HUNTING THE CKM WEAK PHASE WITH TIME-INTEGRATED \\
DALITZ ANALYSES OF $B_s \to K\pi\pi$ DECAYS}

\author{M.~Ciuchini}
\affiliation{Dipartimento di Fisica, Universit\`a di Roma Tre 
and INFN, Sezione di Roma Tre, Via della Vasca Navale 84, I-00146
Roma, Italy}
\author{M.~Pierini}
\affiliation{Department of Physics, University of Wisconsin, Madison,
  WI 53706, USA}
\author{L.~Silvestrini}
\affiliation{Dipartimento di Fisica, Universit\`a di Roma ``La
  Sapienza''  and INFN, 
  Sezione di Roma, P.le A. Moro 2, I-00185 Rome, Italy}

\begin{abstract}
  We present a new technique to extract information on the Unitarity
  Triangle from the study of $B_s \to K\pi\pi$
  Dalitz plot.  Using isospin symmetry and the possibility to access
  the decay amplitudes from Dalitz analyses, we propose a new strategy
  to extract the weak phase $\gamma$ from $B_s \to K\pi\pi$.
\end{abstract}

\maketitle

The present knowledge of the parameters of CP violating and CP
conserving quantities in $B$ and $K$ decays confirms the role of the
Cabibbo-Kobayashi-Maskawa (CKM) matrix as the source of CP violation
in the Standard Model (SM).  The impressive agreement of the available
experimental information in constraining the parameters of the CKM
matrix~\cite{utfit,ckmfit} already reduces the allowed parameter space
of New Physics (NP) contributions to $B_d$ and $K$ decays~\cite{utnp}.
Nevertheless, because of the scarce information available on $B_s$
physics, it is still conceivable to observe large deviations from the
SM in the study of $B_s$ decays~\cite{masiero}.

In this letter, we present a new method to access the CKM phase in
$B_s$ decays, based on a new strategy recently proposed for $B_d \to
K^+ \pi^- \pi^0$ and $B^+ \to K^0 \pi^+ \pi^0$ decays~\cite{cpsBd}.
The study of $B_s \to K^- \pi^+ \pi^0$ is more powerful since it will
directly measure the weak phase $\gamma$ with negligible theoretical
uncertainties and requires the use of isospin symmetry only.  In this
case, in fact, the effect of Electroweak Penguins (EWP's) is marginal,
as we will illustrate in the following.

We stress that the proposed measurements only rely on time integrated
experimental analyses. Therefore our proposal is relevant not only for
hadron collider experiments, but also for the present and future
$B$-factories. Indeed, after Belle clearly proved the possibility of
running a $B$-factory at the $\Upsilon(5S)$ resonance~\cite{belle5S},
measurements of $B_s$ decays can be envisaged both at present
$B$-factories, as the last part of their physics program, and  at a
third-generation high-luminosity linear $B$-factory~\cite{superB}.

Let us first illustrate our idea for the case in which we
neglect EWP contributions. To this aim, we write the amplitudes 
of $B_s \to K^* \pi$ decays using isospin symmetry, in terms of 
Renormalization Group Invariant (RGI) complex parameters~\cite{burassilv},
obtaining
\begin{eqnarray}
  A(B_s \to K^{*-} \pi^+)           &=& ~~V_{tb}^* V_{td} P_1 - V_{ub}^*
  V_{ud} (E_1 - P_1^\mathrm{GIM})  \nonumber\\
  {\sqrt{2}}A(B_s \to \bar K^{*0} \pi^0) &=& -V_{tb}^* V_{td} P_1 - V_{ub}^*
  V_{ud} (E_2 + P_1^\mathrm{GIM}) \nonumber
\end{eqnarray}
where $P_1^\mathrm{(GIM)}$ represent (GIM-suppressed) penguin
contributions, and $E_1$ ($E_2$) the connected (disconnected) emission 
topologies. $\bar B_s$ decay amplitudes are simply obtained by conjugating 
the CKM factors $V_{ij}$. Similar expressions hold for higher $K^*$ resonances.

Considering the two combinations of amplitudes
\begin{eqnarray}
A_s^{K^*\pi}&=&A(B_s \to K^{*-}\pi^+)+{\sqrt{2}}A(B_s \to \bar K^{*0} \pi^0)\nonumber\\
&=&-V_{ub}^* V_{ud} (E_1 +E_2)\,,\nonumber\\
\bar A_s^{K^*\pi}&=&A(\bar B_s \to K^{*+}\pi^-)+{\sqrt{2}}A(\bar B_s \to K^{*0} \pi^0)\nonumber\\
&=&-V_{ub} V_{ud}^* (E_1 +E_2)\,,\label{eq:atnb}
\end{eqnarray}
the ratio
\begin{equation}\label{eq:rn}
R_d=\frac{\bar A_s^{K^*\pi}}{A_s^{K^*\pi}}=\frac{V_{ub} V_{ud}^*}{V_{ub}^* V_{ud}}=e^{-2 i\gamma}
\end{equation}
provides a clean determination of the weak phase $\gamma$.  Since the
final states are self-tagging, the phase of the $B_s$--$\bar B_s$
mixing amplitude does not enter our analysis.

The $A_s^{K^*\pi}$ amplitudes can be experimentally determined looking
at the decay chains $B_s \to K^{*-}(\to K^-\pi^0)\pi^+$ and $B_s \to
\bar K^{*0}(\to K^-\pi^+)\pi^0$ in the $B_s \to K^-\pi^+\pi^0$ Dalitz plot.
Similarly, $\bar A_s^{K^*\pi}$ can be extracted from the
$K^+\pi^-\pi^0$ Dalitz plot. However, one has to fix the phase
difference between $A(K^{*-}\pi^+)$ and $A(K^{*+}\pi^-)$. 

This information can be provided by the $K_s\pi^+\pi^-$ Dalitz plot,
considering the decay chain $B_s \to K^{*-}(\to \bar K^0\pi^-)\pi^+$
and the CP conjugate $\bar B_s \to K^{*+}(\to K^0\pi^+)\pi^-$. These
two decay channels do not interfere directly on the Dalitz plot, but
they both interfere with the decays $B_s,\bar B_s\to \rho^0 (\to
\pi^+\pi^-) K_s$, $B_s,\bar B_s\to f^0 (\to \pi^+\pi^-) K_s$ and with
all other $\pi^+\pi^-$ resonant and possibly non-resonant intermediate
states. Therefore the Dalitz analysis of $B_s,\bar B_s\to
K_s\pi^+\pi^-$ should include these states. In more details, one needs
to measure at least four amplitudes: $A(B_s\to K^{*-}\pi^+)$,
$A(B_s\to \rho^0 K_s)$, $A({\bar B}_s\to K^{*+}\pi^-)$, and $A({\bar
  B}_s\to \rho^0 K_s)$. The interference between $A(B_s\to
K^{*-}\pi^+)$ and $A(B_s\to \rho^0 K_s)$ fixes the relative phase of
these two amplitudes. The same applies to $A({\bar B}_s\to
K^{*+}\pi^-)$ and $A({\bar B}_s\to \rho^0 K_s)$. The only missing
ingredient to be determined is the relative phase between $A(B_s\to
\rho^0 K_s)$ and $A({\bar B}_s\to \rho^0 K_s)$. The extraction of this
phase depends on the experimental environment. In the untagged case,
the cosine of the phase difference can be extracted from
time-integrated measurements through the interference term
proportional to
$y=\Delta\Gamma_s/2\Gamma_s$~\cite{Bigi:1981qs,Gronau:2006qn}. In the
correlated case at the $\Upsilon(5s)$, more information is accessible
through a tagged time-integrated analysis, provided positive and
negative time differences between the tag and decay sides can be
distinguished. Taking the difference of positive and negative
time-integrated rates, one can extract also the sine of the phase
difference through the interference term proportional to
$x/(1+x^2)\sim \Gamma_s/\Delta m_s$.

As discussed for the case of $B_d$ decays~\cite{cpsBd}, one should consider
the inclusion of EWP's. In the case of $R_d$, the effect of EWP's is
very small since the $\mathcal{O}(\alpha_{em})$ suppression is not
compensated by any Cabibbo enhancement, so that eq.~(\ref{eq:rn})
remains valid with an uncertainty of $\sim 3 \%$. Let us now illustrate this
error estimate. The dominant EWP's
(\textit{i.e.}~left-handed EWP operators) can be eliminated at the
operator level.  Following the notation and derivation in
Ref.~\cite{cpsBd}, we can write the effective Hamiltonian for $b \to
d$ transitions as
\begin{eqnarray}\label{eq:heffapp}
  H_\mathrm{eff} &\simeq& \frac{G_F}{\sqrt{2}}\Biggl\{V_{ub}^* V_{ud} \left(1+ \kappa_\mathrm{\scriptscriptstyle EW}\right)
  \Bigl[C_+ \left(Q_+^{duu} - Q_+^{dcc}\right)+ \\ && \quad
  \frac{1-\kappa_\mathrm{\scriptscriptstyle EW}}{1+\kappa_\mathrm{\scriptscriptstyle EW}}
  C_-\left(Q_-^{duu} - Q_-^{dcc}\right)\Bigr]-
  V_{tb}^* V_{td} H^{\Delta  I=1/2}\Biggr\}\nonumber
\end{eqnarray}
with
\begin{eqnarray}
  \kappa_\mathrm{\scriptscriptstyle EW}&\equiv& - \frac{3}{2} \frac{C_+^\mathrm{\scriptscriptstyle EW}}{C_+}
  \frac{V_{tb}^* V_{td}}{V_{ub}^* V_{ud}}\,,
  \label{eq:kappa}
\end{eqnarray}
where $C_+^\mathrm{\scriptscriptstyle (EW)}$ are defined in
Ref.~\cite{cpsBd}.  The correction term
$\kappa_\mathrm{\scriptscriptstyle EW}$ can be neglected. For example,
using the values $C_+(m_b) = 0.877$, $C_+^\mathrm{\scriptscriptstyle
  EW}(m_b) = -1.017 \, \alpha_{em}$ and $\bar\rho = 0.216$, $\bar\eta
= 0.342$~\cite{utfit}, we obtain $\kappa_\mathrm{\scriptscriptstyle
  EW}=(0.4+2.8\, i)\times 10^{-2}$, so that, as we anticipated, EWP's
are negligible in $R_d$~\footnote{As discussed in
  Ref.~\cite{Gronau:2006qn}, the $\Delta I=3/2$ matrix element of
  $Q_-$ vanishes in the exact isospin symmetric limit allowing for an
  exact treatment of the EWP's.  However the uncertainty associated to
  isospin breaking is of the same order of our estimate.}.

Let us finally comment on the sensitivity to NP of our analysis.
$R_d$ is unaffected by any NP entering at the loop level (barring
order-of-magnitude enhancements of EWP's) so that it is on the same
footing as other measurements of gamma from pure tree decays.

We have presented a new method to constrain the Unitarity Triangle,
starting from $B_s \to K \pi \pi$ decays, using the amplitude ratio
$R_d$ defined in Eq.~(\ref{eq:rn}). The method determines the weak
phase $\gamma$, up to a negligibly small correction coming from EWP
operators. In addition, the ratio $R_d$ is unaffected by NP entering
at the loop level. Since these measurements do not require
time-dependent studies, they offer the possibility of accessing the
CKM phase $\gamma$ both at hadron colliders and at high-luminosity
$B$-factories.

This work has been supported in part by the EU networks ``The quest
for unification'' under the contract MRTN-CT-2004-503369 and
``FLAVIAnet'' under the contract MRTN-CT-2006-035482.

{\bf Note added.}~During the revision of this paper,
Ref.~\cite{Gronau:2006qn} appeared where our results were confirmed
and further arguments were given to control EWP effects.

\end{document}